# Structural Phase Transitions and Sodium Ordering in $Na_{0.5}CoO_2$: a Combined Electron Diffraction and Raman Spectroscopy Study

H.X. Yang*, C.J. Nie, Y.G. Shi, H.C. Yu, S. Ding, Y.L. Liu, D. Wu, N.L. Wang and J.Q. Li

Beijing National Laboratory for Condensed Matter Physics, Institute of Physics, Chinese Academy of Sciences, Beijing 100080, China

Author to whom correspondence should be addressed: hxyang@blem.ac.cn

# Abstract

The nonstoichiometric $Na_xCoO_2$ system exhibits extraordinary physical properties that correlate with temperature and Na concentration in its layered lattice without evident long-range structure modification when conventional crystallographic techniques are applied. For instance, $Na_{0.7}CoO_2$, a thermodynamically stable phase, shows large thermoelectric power; water-intercalated $Na_{0.33}CoO_2 \cdot 1.3H_2O$ is a newly discovered superconductor with $T_c \sim 4K$, and $Na_{0.5}CoO_2$ exhibits an unexpected charge ordering transition at around $T_{co} \sim 55$ K. Recent studies suggest that the transport and magnetic properties in the $Na_xCoO_2$ system strongly depend on the charge carrier density and local structural properties. Here we report a combined variable temperature transmission electron microscopy and Raman scattering investigation on structural transformations in $Na_{0.5}CoO_2$ single crystals. A series of structural phase transitions in the temperature range from 80 K to 1000 K are directly identified and the observed superstructures and modulated phases can be interpreted by Na-ordering. The Raman scattering measurements reveal phase separation and a systematic evolution of active modes along with phase transitions. Our work demonstrates that the high mobility and ordering of sodium cations among the $CoO_2$ layers are a key factor for the presence of complex structural properties in $Na_xCoO_2$ materials, and also demonstrate that the combination of electron diffraction and Raman spectroscopy measurements is an efficient way for studying the cation ordering and phase transitions in related systems.

# 1. Introduction

Layered deintercalatable alkali metal oxides, such as $Li_xCoO_2$ and $Na_xCoO_2$, have been a subject of an intense research activity in the past years owing to their potential technological applications as a battery electrodes and thermoelectric materials [1-11]. The notable features of these materials, from both structural and chemical point of views, are that the cation content and crystal structure can vary over a wide range by deintercalation, and that the structural and physical properties are profoundly affected by the cation concentration and cation and vacancy ordering [4-5, 11]. Experimental measurements, however, indicate that it is very difficult to obtain comprehensive structural information using crystallographic techniques, owing to considerable positional disorder of the intercalated cations and phase segregations, as identified in our recent investigations [12]. It is therefore necessary to perform systematic investigations using a variety of methods to gain insight into the local atomic arrangements, especially regarding the intercalated cations between the metal oxides layers.

Recently, extensive interest has been paid to structure and physical properties of $Na_{0.5}CoO_2$ which shows three low-temperature phase transitions at around 87 K, 53 K, and 25 K respectively. These phase transformations, possibly in connection with charge ordering, have been well identified by the measurements of resistivity, thermal transport and magnetization. The 87 K transition arises partially from a structural change; the 53 K transition is found to be responsible for charge ordering associated with a magnetic ordering; and the 20 K transition is proposed to be a spin reorientation transformation [13].

Neutron diffraction at ~9K identified an orthorhombic structure for $Na_{0.5}CoO_2$ with the Na atoms completely ordered [13]. TEM experimental investigations reveal that complicated Na-ordered states and phase segregations exist commonly in $Na_xCoO_2$ due to existence of two crystallographic positions and high mobility of $Na^+$ [12]. It is well known that TEM is ideal for the study of crystal structure at the nano-scale and that Raman scattering is sensitive to the local atomic arrangement change induced by oxygen shifts and $Na^+$ ordering. In this paper, a detailed Raman scattering study, revealing the changes of lattice vibrations with temperature, and TEM observations, revealing structural transitions, have been carried out to study the structural properties in $Na_{0.5}CoO_2$ material. A series of structural phase transitions, as observed in the temperature range from 70 K up to 1000 K has been analyzed in terms of $Na^+$ ordering.

## 2. Experimental

Single crystalline $Na_{0.85}CoO_2$ samples were grown using a traveling-solvent floating zone method, $Na_{0.5}CoO_2$ compounds were further prepared by sodium deintercalation of $Na_{0.85}CoO_2$ as describe in previous publications. The sodium content of all samples was determined by the ICP method. Specimens for transmission-electron microscopy (TEM) observations were prepared simply by crushing the bulk material into fine fragments, which were then supported on a copper grid coated with a thin carbon film. The TEM investigations were performed on an H-9000NA TEM operating at a voltage of 300 kV, and a TECNAI 20 operating at a voltage of 200 kV, both equipped with low and high temperature sample stages. Raman spectra were collected in back-scattering geometry from 470 K down to 79 K using a Jobin-Yvon T64000 triple spectrometer equipped with a cooled change-couple device. In the spectrometer an objective of 100X-magnification was used to focus the laser beam on the sample surface and to collect the scattered light. Two excitation wavelengths at 488.0nm and 514.5 nm of an $Ar^+$ ion laser were used in our experiments. The laser power at the focus spot of 2-3 μm in diameter was kept below 1 mW to prevent the samples from laser-induce damage during experiments.

## 3. Results and discussion

We first consider the microstructure changes in $Na_{0.5}CoO_2$ and development of local $Na^+$ order with temperature. We have performed *in situ* heating TEM studies from 300 K to 1000 K and *in situ* cooling from 300 K to 100 K. A series of phase transitions related to $Na^+$ order have been identified at different temperatures. The material has an incommensurate modulated structure at room temperature and upon in-situ heating, it transforms into a superstructure hexagonal phase with doubled cell parameters in *a-b* plane: this superstructure is stable in the temperature range between 410 K and 470 K where upon another structural transition appears at about 470 K and the $Na_{0.5}CoO_2$ material transforms into a high-temperature hexagonal structure. Upon cooling, $Na_{0.5}CoO_2$ undergoes a transformation at ~200 K towards an orthorhombic structure [13, 18] with space group *Pnmm* in which a charge-ordered state is observed at the lower temperature. Figs. 1 (a)-(d) show the typical selected-area electron diffraction patterns taken at different temperatures demonstrating the presence of a series of remarkable phase transitions in the $Na_{0.5}CoO_2$ material.

We now proceed to a detailed discussion of the structures observed at different temperatures. TEM structural analyses suggest that the high temperature hexagonal phase (above 470K) has the similar average structure to the conventional hexagonal phase in which no any ordered states are observed (see Fig.1 (a)). This ideal (average) structural homogeneity is perhaps in connection with the high mobility of Na ions at high temperatures. Moreover, the thin TEM samples are found to gradually decompose into small polycrystalline particles above 850 K.

The notable superstructure appearing between 410 K and 470 K (see Fig.1 (b)) has doubled cell parameters within the *a-b* plane and this superstructure is interpreted as arising from partial $Na^+$ ordering between $CoO_2$ layers. The $Na^+$ ions in reported average structures can occupy two possible positions ($Na_1$ and $Na_2$) governed by the space between the two adjacent O planes. Both of the sites are trigonal prismatic sites, but one trigonal prism shares edges with adjacent $CoO_6$ octahedral ($Na_2$), while the other trigonal prism shares faces with adjacent $CoO_6$ octahedral ($Na_1$) [13]. Fig. 2 (a) displays a simple structural model based on the symmetric occupation of a fraction of the $Na_2$ sites. Fig. 2 (b) shows a theoretical simulated diffraction pattern along the [001] zone axis direction, which is in good agreement with the experimental results as exhibited in Fig.1 (b). It is also noted that $Na_{0.5}CoO_2$ in this temperature range also has perfect structural homogeneity. We have checked the crystal structure from one area to another and no other kind of additional ordering or structural changes are observed.

Structural data obtained between 200 K and 410 K, especially close to room temperature, reveal a variety of interesting structure phenomena, such as phase separation and complex ordered states corresponding to different Na-ordered states [12]. It is likely that $Na_{0.5}CoO_2$ has an intermediate (or metastable) structure within this large temperature range. Detailed TEM structural analyses suggest that the major structural feature of this intermediate phase can be described by a structural modulation along the <110> direction. This modulation in certain areas is extremely weak yielding the conventional hexagonal structure [12]. In some other areas, however, it is apparent as a commensurate superstructure and gives rise to an orthorhombic structure. Fig. 1 (c) represents a [001] zone-axis diffraction pattern that exhibits a clear incommensurate structural modulation approximately along the <110> direction. All satellite spots following the main spots in the pattern can be indexed by a structural modulation with a modulation wave vector q=<1/4, 1/4, 0>+δ, where the incommensurate parameter δ, changing slightly from area to area, ranges from 0 to 1/24 and depends on temperature. Fig. 3 (a) displays the electron diffraction patterns for three different structural modulations at room temperature. The top diffraction pattern is the most frequently observed one, showing both orientational and spatial anomalies as indicated by the dashed lines. The middle diffraction pattern shows a visible incommensurability with δ~1/16 without orientational anomaly and the bottom diffraction pattern, corresponding to an ideal superstructure (δ=0), appears in a few areas at room temperature. It should be mentioned that although the modulated structure appears in a large fraction of areas, a notable fraction of samples (~30%) still have the well-defined hexagonal average structure where this structural modulation is almost invisible.

According to theoretical results for structural modulations, the incommensurate state consists of small commensurate domains separated by discommensurations within which the phase of structural modulation varies rapidly [17]. Fig. 3 (b) shows a dark-field TEM image revealing the presence of a lamella structural feature where the average periodicity changes slightly from area to another (~ 5nm). This kind of domain structure is often seen in incommensurate or nearly commensurate modulated systems. The white stripes are considered as commensurate regions where the modulation phase factor remains constant. The average periodicity ($L$) of this type of domain structure is related to the incommensuratability $\delta$ through the relationship $L = 2\pi/p\delta$, where $p$ is the order of commensurability and, in the present case, $p = 4$. In $Na_{0.5}CoO_2$, we have $\delta=1/15$, this gives rise to a periodical domain structure with L ~5 nm. This result is in good agreement with the periodicity observed in the dark field image.

The mechanism for $Na^+$ ordering over the available sites has been discussed previosuly using the model that the Na atoms are spaced as far apart as possible to minimize the $Na^+$- $Na^+$ Coulombic repulsions [13]. Occupancy of the $Na_2$ site is favored as demonstrated by neutron diffraction for $Na_{0.61}CoO_2$ [19]. A structural model for explaining the incommensurate modulated structure is reported in ref. 18, in which only one of the two $Na_1$ and $Na_2$ sites is partially occupied. The incommensurability is proposed to originate from an insertion of an extra vacancy plane. Fig. 3 (c) and (d) display two structural models which could be used to explain the two incommensurate modulations in Fig 3 (a). These models give a reasonable fit with the observed electron diffraction patterns, and both orientation and space anomalies can be explained.

Now we move on to discuss the structure properties of the low-temperature structural phase transition. TEM observations reveal that the incommensurate structural modulation shows up as systematic changes in wavelength, intensity and incommensurability with the decrease of temperature. The satellite reflections at temperatures above 200 K in general emerge as weak spots and under *in situ* cooling their intensity increases rapidly at the temperature just below 200 K. The incommensurate structural modulation undergoes a lock-in phase transition to a well-defined orthorhombic structure as demonstrated by the diffraction observations (see Fig. 1 (d)). A structural model for this phase is proposed with Na ordering [13], in which one out of four $Na_1$ as well as $Na_2$ sites are occupied and zig-zag Na chains are formed between two neighboring $CoO_2$ layers. Figs. 3 (e) and (f) qualitatively illustrate the intensity and the incommensurability of modulation, *q*, as a function of the temperature obtained from several single crystal samples. These results directly suggest that the structural changes occur just below 200 K for the $Na_{0.5}CoO_2$ system.

Raman scattering measurements on single crystalline samples of $Na_{0.5}CoO_2$ indicate that structural inhomogeneity commonly appears at room temperature and this result is consistent with our previous TEM data that showed the presence of phase separation in $Na_{0.5}CoO_2$ materials [12]. Fig. 4 (a) shows three typical Raman spectra obtained from different areas on a single crystal at room temperature. The focus spot of the laser beam in our measurements is about 2-3 μm in diameter. Notable distinctions among these Raman spectra from different areas can be clearly recognized in both peak positions and intensities: these data directly demonstrate the presence of structural inhomogeneity (phase separation) in the γ-$Na_{0.5}CoO_2$ single crystals. The top spectrum in Fig. 4 (a), containing five clear Raman peaks, can be easily matched to the hexagonal average structure as reported in our previous paper [20]. These five peaks are assigned respectively as $A_{1g}$ at 681cm$^{-1}$, $E_{1g}$ at 193cm$^{-1}$, $E_{2g}$ at 476cm$^{-1}$, 520cm$^{-1}$, and 614 cm$^{-1}$ for the space group of $P6_3/mmc$. The bottom spectrum in Fig. 4 (a) shows very different Raman active modes from that of the hexagonal phase, and contains three distinct phonon modes at 439, 476, and 573 cm$^{-1}$. Careful analysis (see below) suggests that this spectrum is taken from an area dominated by the presence of an orthorhombic phase. The middle spectrum in Fig. 4 (a) shows recognizable features arising from the combination of the other two spectra. This last of spectrum is frequently observed in our experiments and is considered to originate from structural inhomogeneities characterized by an incommensurate structural modulation along the <110> direction. It is also noted that besides the extra peak around 573 cm$^{-1}$, possibly from orthorhombic structural distortion, the five Raman peaks corresponding to the hexagonal crystal lattice shift towards the higher frequencies in comparison with that in the top spectrum of hexagonal phase. The $A_{1g}$ mode, in particular, changes evidently in both position and intensity.

In accordance with previous study of the $Na_xCoO_2$ hexagonal average structure [20-21], the $A_{1g}$ and $E_{1g}$ modes only involve motions from oxygen atoms; the $E_{1g}$ mode is an in-plane oxygen mode with diagonal displacements while the $A_{1g}$ mode is an out-of-plane mode. The energy of $A_{1g}$ mode strongly depends on the Na site occupancy and on the Na occupation in the layer that divides the oxygen octahedral in the c-axis direction, Hence, the $A_{1g}$ mode is a sensitive factor in the analysis of Na distribution/ordering in $Na_xCoO_2$ system.

In order to further understand the variation of the Raman active modes in relation to the structural transitions in the high-temperature range, we have carried out *in situ* heating Raman scattering experiments on several single crystal samples. The high temperature experimental results demonstrate that all three types of spectra in Fig. 4 (a) undergo certain transitions toward a common spectrum similar to that of the hexagonal phase. Fig. 4 (b) shows the corresponding spectra of Fig. 4 (a) taken at temperatures of around 450 K. The spectrum taken from hexagonal phase (area A) only shows limited changes with temperature rise; on the other hand, the spectrum from the orthorhombic phase (area C) evolutes progressively into a spectrum similar to the hexagonal structure with increase of temperature. Experimental measurements indicated that rapid spectrum changes occur at around 390 K to 420 K. It is also noted that three high temperature spectra still contain noticeable differences, indicating the presence of local structural inhomogeneities in the areas studied.

We have also performed *in situ* cooling Raman scattering measurements to reveal changes of active modes with the hexagonal-orthorhombic phase transition. Fig. 5 shows a series of Raman spectra starting from the hexagonal structure at room temperature and ending at the orthorhombic phase at ~80 K. The peak that appears at 573 cm$^{-1}$, which likely represents the growth of orthorhombic phase, develops at around 200 K. From 200 K down to 140 K, the peaks at 193 cm$^{-1}$ and 681 cm$^{-1}$ of the hexagonal structure decrease in intensity quickly. These rapid changes in the lattice clearly demonstrate the phase transition starting from 200K from the hexagonal to orthorhombic phase as discussed above. A further analysis (as shown in Fig. 5 (b)) illuminates that, from 300 K to 200 K, the $E_{2g}$ mode at 474 cm$^{-1}$ and the $A_{1g}$ mode at 681cm$^{-1}$ shift progressively to higher frequencies.

Moreover, the line width of the $A_{1g}$ mode becomes substantially sharper. These transformations can be considered to be correlated with confined sodium motion and ordering with lowering temperature.

Based on our structural investigations, the low temperature spectrum (80K) should correspond with the well-defined orthorhombic phase. From classical factor group analysis for the *Pmmn* space group, there are 24 Raman active phonon modes described as following: $7A_g+4B_{1g}+7B_{2g}+6B_{3g}$, in sharp contrast with five modes $A_{1g} + E_{1g} + 3E_{2g}$ for hexagonal phase. Our observations reveal only three or four visible peaks in the experimental spectra of measured range as shown in Fig. 5. Detailed analysis on the low temperature orthorhombic phase, especially spectrum change along with charge order transition in present system, is still in progress.

Numerous unusual experimental phenomena in $Na_xCoO_2$ materials have been discussed in connection with the sodium ordering in the recent literature [14-16]. For instance, infrared spectroscopy measurements on $Na_{0.5}CoO_2$ crystals show dramatic changes in the phonon modes above 83 K [15]; dielectric investigations on $Na_{0.85}CoO_2$ demonstrate significant phonon modes changes between 280 and 260 K [14] and the $^{23}$Na-NMR spectra study show only small changes of the Larmor frequencies of Na nuclei in the temperature range of 240 - 260 K [16]. These observations have been partially interpreted as the evidence of the motion changes of Na ions in the low temperature range.

## 4. Conclusions

In summary, the material $Na_{0.5}CoO_2$ has a rich variety of structural phases at different temperatures and the fundamental structural features in these phases can be understood as arising from Na ordering, which strongly modulates the conventional hexagonal layered structure. It is remarkable that between 210 K and 410 K, the coexistence and phase competition between the hexagonal and orthorhombic structures yield phase segregation as revealed by TEM observations and Raman scattering measurements. Between 410 K and 47 K, $Na_{0.5}CoO_2$ has a hexagonal superstructure with doubled cell parameters in the *a-b* plane, and below 210 K, $Na_{0.5}CoO_2$ has an orthorhombic structure in which charge and magnetic ordered states are identified at about 55 K. Our combined electron diffraction and Raman spectroscopy study is demonstrated to be an effective way in which to reveal the structural properties arising from Na cation ordering, phase separation and phase transitions in the $Na_xCoO_2$ system.

## Acknowledgments

We would like to thank Professor N.L. Wang for providing single crystals of $Na_xCoO_2$ and Miss G. Zhu for the assistance in preparing samples and measuring Raman spectra. Thanks Dr R.I. Walton for his help during manuscript preparation. The work reported here is supported by National Natural Foundation of China.

# Figure captions

Fig.1 Electron diffraction patterns showing the structural phase transitions in $Na_{0.5}CoO_2$, taken at the temperatures of at (a) 500 K, (b) 420 K, (c) 300 K, and (d) 100 K.

Fig. 2 (a) Structural model for the superstructured phase observed between 410 K and 470 K. the occupied Na sites are indicated by an arrow. (b) The simulated electron diffraction pattern along the [001] zone axis direction.

Fig. 3 (a) Electron diffraction pattern showing three kind of structural modulations in $Na_{0.5}CoO_2$ at room temperature. (b) Dark field TEM image from a satellite spot showing the lamella structure in connection with incommensurate modulation. Structural models for the incommensurate modulations (c) with and (d) without orientation anomaly, only Co sites and occupied Na sites are shown. Temperature dependences of (e) intensity and (f) periodicity of structural modulation in $Na_{0.5}CoO_2$.

Fig. 4 Raman spectra taken from three different areas in a $Na_{0.5}CoO_2$ single crystal (a) at room temperature, showing three types of spectra from phase separation, and (b) at 450 K, illustrating evident evolutions in both spectra from area B and area C.

Fig. 5 (a) A series of Raman spectra from room-temperature hexagonal structure down to low- temperature orthorhombic structure (80 K), demonstrating a systematic change in both peak position and intensity. (b) Detailed analysis of spectrum evolution from 300 K down to 200 K.

.

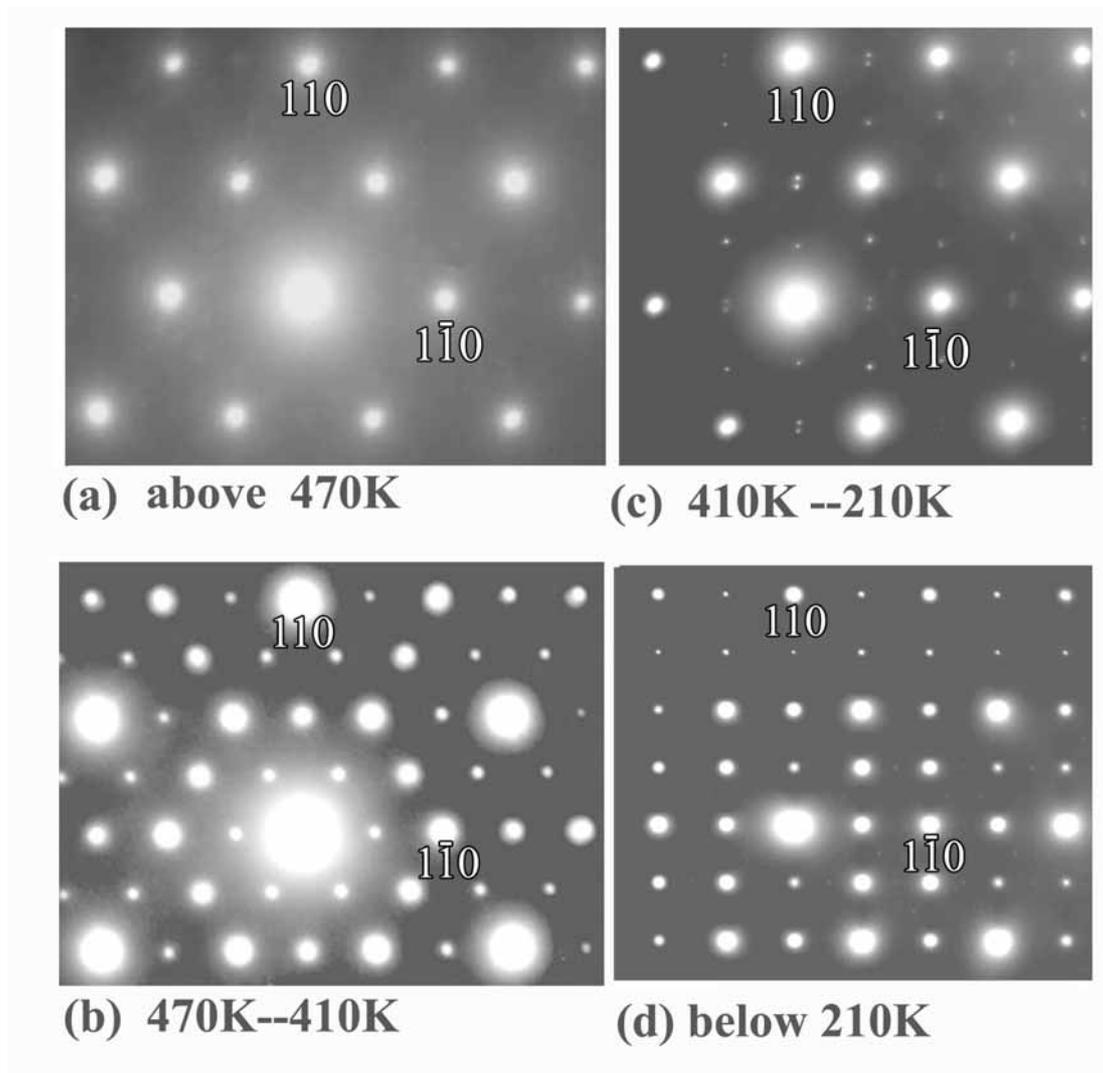

Yang et al. (Fig. 1)

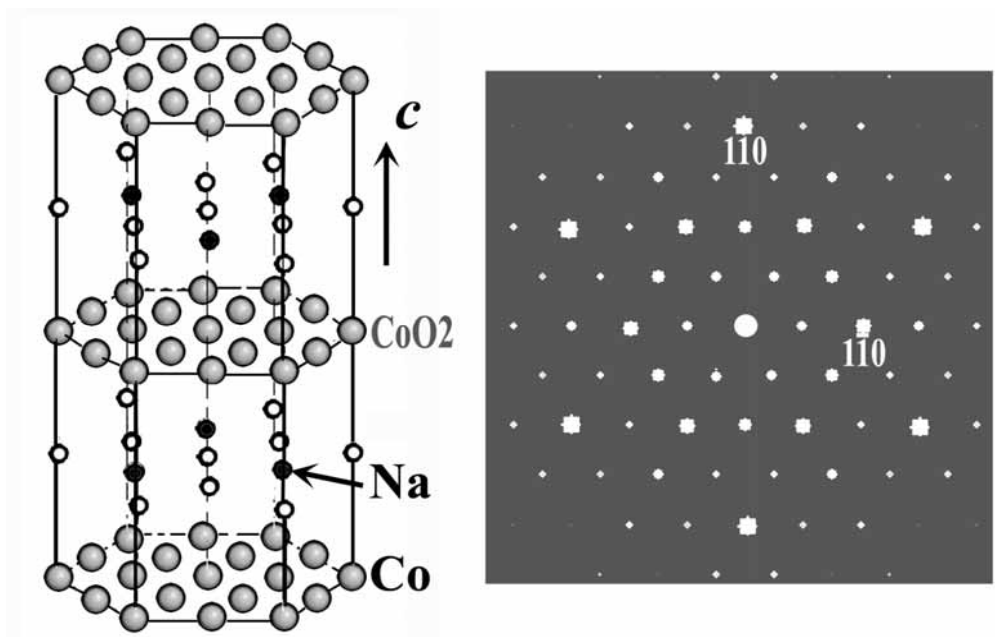

(a)　　　　　　　　　　　　　　　(b)

Yang et al. (Fig. 2)

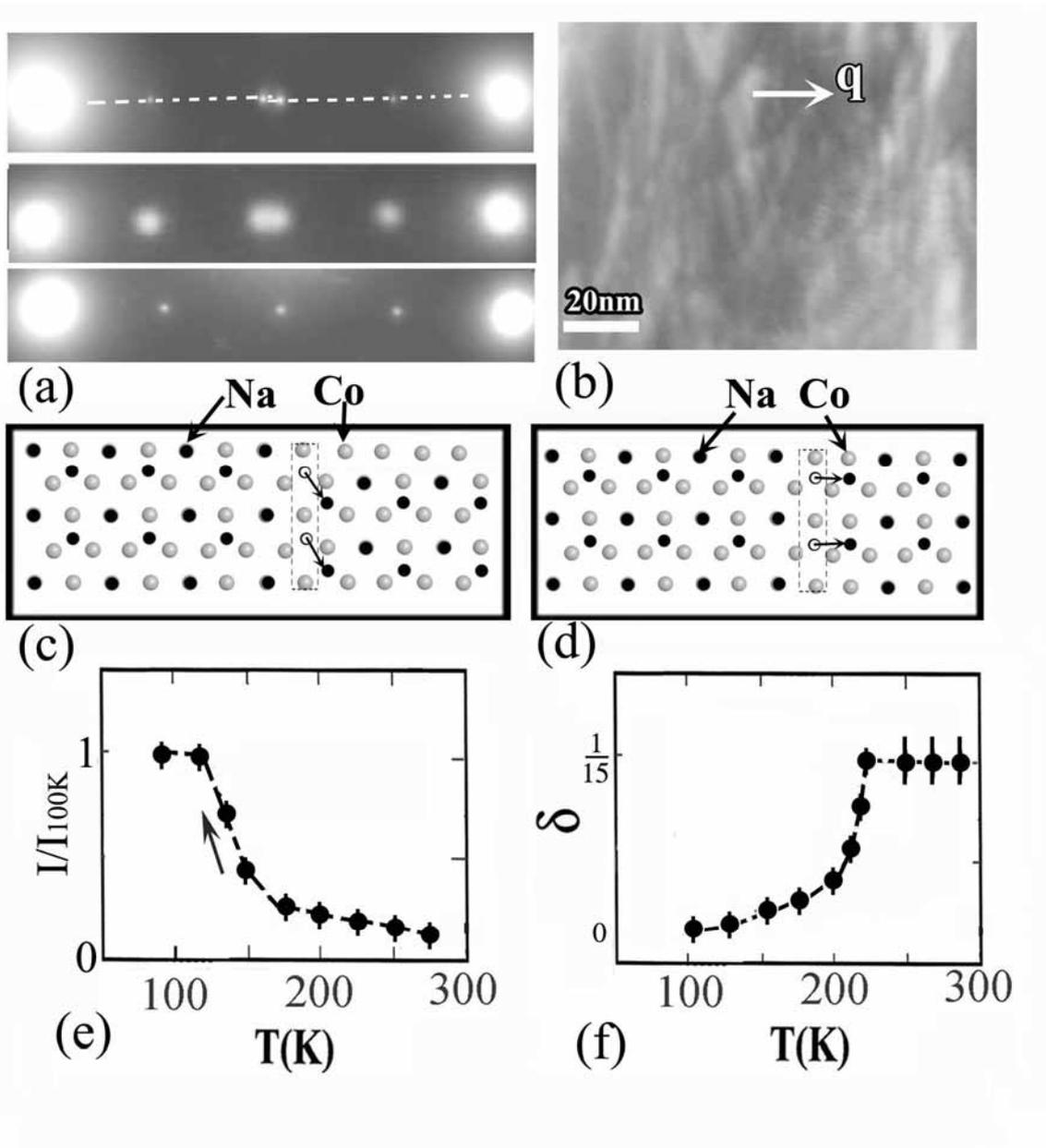

Yang et al. (Fig. 3)

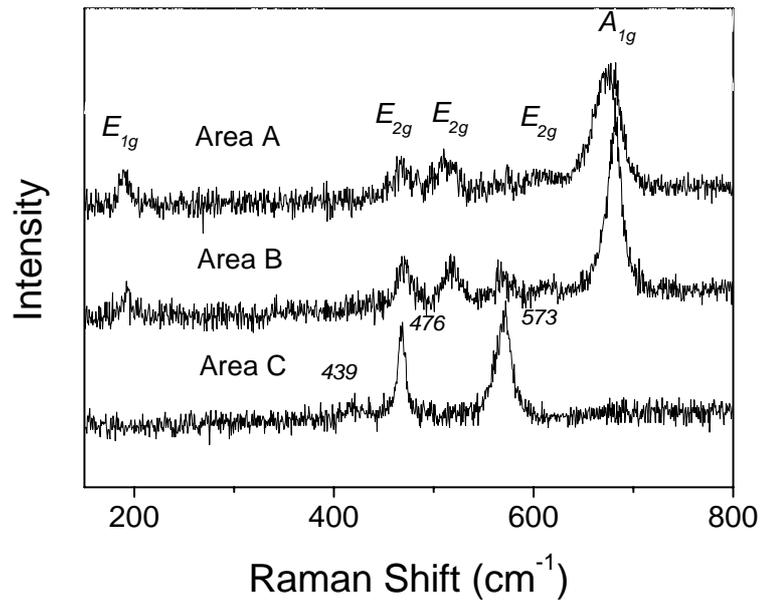

(a)

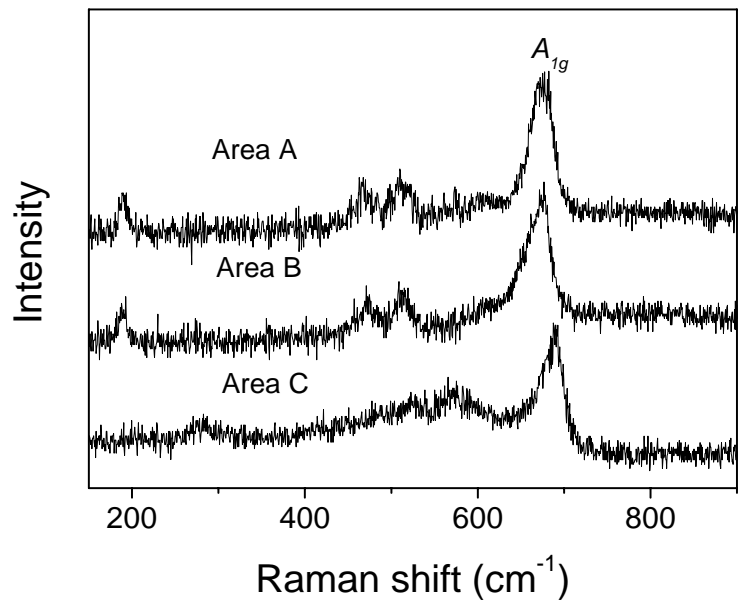

(b)

Yang et al. (Fig. 4)

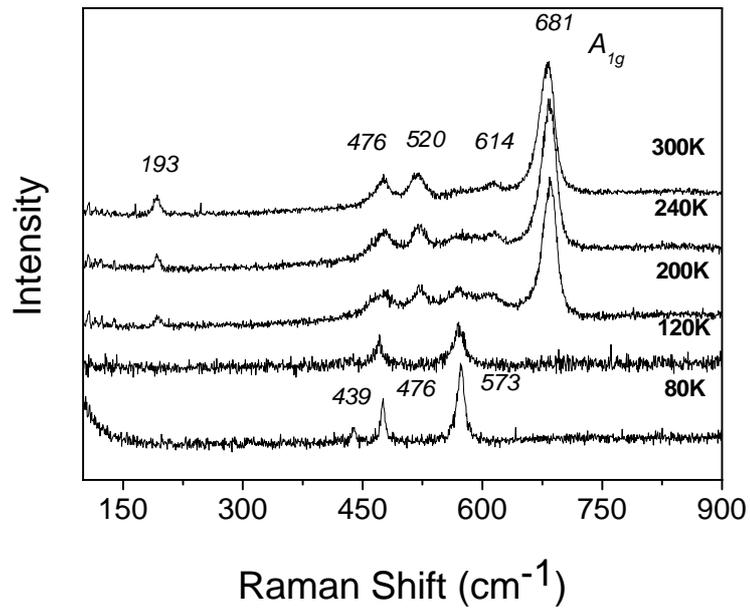

(a)

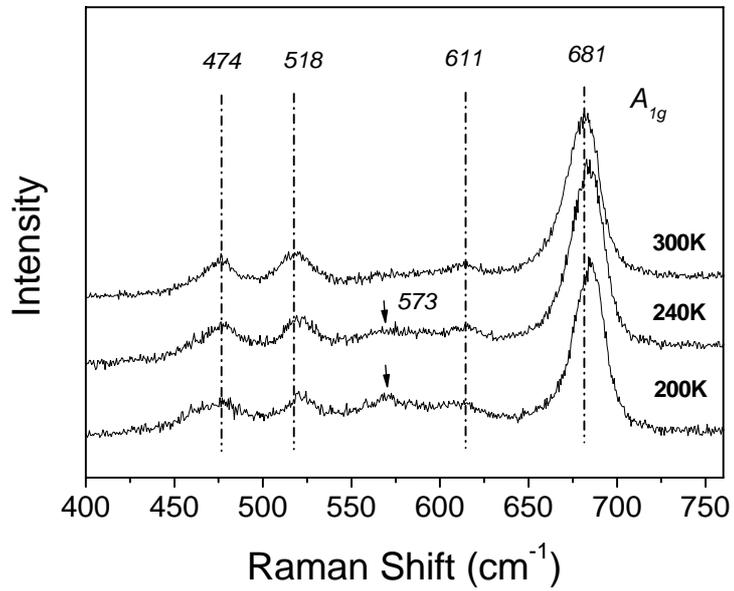

(b)

Yang et al. (Fig. 5)